\documentclass[conference]{IEEEtran}
\IEEEoverridecommandlockouts
\usepackage[dvipsnames]{xcolor}
\usepackage{cite}
\usepackage{amsmath,amssymb,amsfonts}
\usepackage{float}
\usepackage{algorithmic}
\usepackage{graphicx}
\usepackage{textcomp}
\usepackage[ruled,vlined]{algorithm2e}
\usepackage{xcolor}
\usepackage[pagebackref=false,breaklinks=true,colorlinks,bookmarks=false,linkcolor=black,urlcolor=black,citecolor=black]{hyperref}
\def\BibTeX{{\rm B\kern-.05em{\sc i\kern-.025em b}\kern-.08em
    T\kern-.1667em\lower.7ex\hbox{E}\kern-.125emX}}
\begin{document}

\newcommand{\ZL}[1]{%
{#1}}
\newcommand{\hm}[1]{%
{#1}}
\newcommand{\pl}[1]{%
{#1}}

\title{Quantifying the Impact of Lossy Compression on Neural Generative Surrogate
Modeling\\
{}
\thanks{%
LLNL-CONF-2007282}
}

\makeatletter 
\newcommand{\linebreakand}{%
  \end{@IEEEauthorhalign}
  \hfill\mbox{}\par
  \mbox{}\hfill\begin{@IEEEauthorhalign}
}
\makeatother 

\author{\IEEEauthorblockN{Zhimin Li}
\IEEEauthorblockA{\textit{Department of Computer Science} \\
\textit{Vanderbilt University}\\
Nashville, United States \\
zhimin.li@vanderbilt.edu}
\and
\IEEEauthorblockN{Harshitha Menon}
\IEEEauthorblockA{\textit{Center for Applied Scientific Computing} \\
\textit{Lawrence Livermore National Laboratory}\\
Livermore, United States \\
harshitha@llnl.gov}
\and
\IEEEauthorblockN{Charles Jekel}
\IEEEauthorblockA{\textit{Computational Engineering Division} \\
\textit{Lawrence Livermore National Laboratory}\\
Livermore, United States \\
jekel1@llnl.gov}
\and
\linebreakand
\IEEEauthorblockN{Valerio Pascucci}
\IEEEauthorblockA{\textit{School of Computing} \\
\textit{
University of Utah}\\
Salt Lake City, United States \\
pascucci@sci.utah.edu}
\and
\IEEEauthorblockN{Peter Lindstrom}
\IEEEauthorblockA{\textit{Center for Applied Scientific Computing} \\
\textit{Lawrence Livermore National Laboratory}\\
Livermore, United States \\
pl@llnl.gov}
}

\maketitle

\begin{abstract}
Neural networks are used as generative surrogate models for scientific discovery, which are trainable approximations of scientific simulations. These models enable users to replace time-consuming numerical simulations with learned alternatives, providing quick solutions. However, high-fidelity generative surrogate models require massive training datasets, which can create storage and I/O challenges. Lossy compression is a promising way to reduce this burden, but compression errors may affect the model quality in subtle ways, making it challenging to quantify their impact.
   
In this work, we examine how lossy compression of training data impacts the quality of generative surrogate models.
We begin by characterizing the uncertainty inherent in training neural networks, showing that identical training configurations can produce different models. 
By exploiting this variability, we propose a method to estimate how much compression-induced error a surrogate model can tolerate without affecting its accuracy. 
Evaluation of two application simulations demonstrates that our approach significantly reduces memory/storage requirements and speeds up training while producing high-quality surrogate models. 
These results show that lossy compression saves data storage up to 23.7$\times$ and 39$\times$ with negligible impact on the quality of the surrogate model. Meanwhile, reducing the size of the training data set also enhances the data loading speed and reduces the training time by up to 3$\times$.

\end{abstract}

\begin{figure*}[!htbp]
\centering   
    \includegraphics[width=\linewidth]{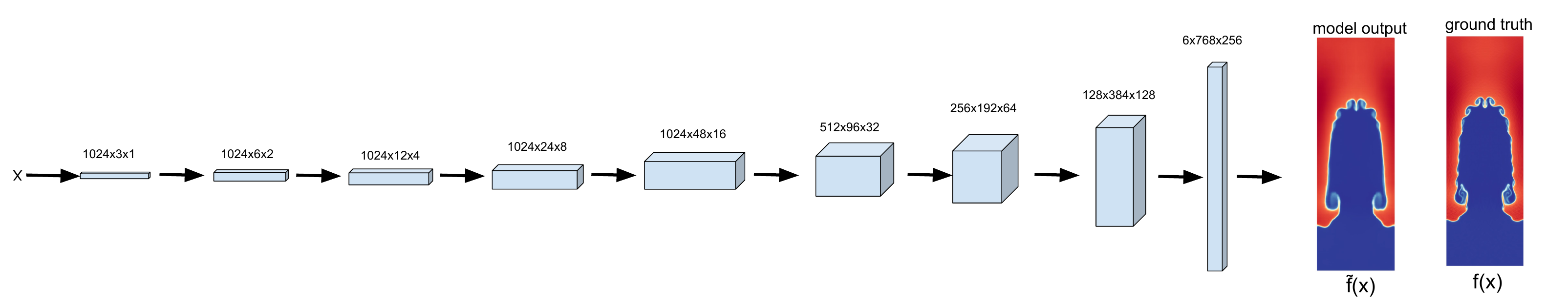}
    \caption{The architecture used for neural surrogate model training and simulation generation.}
\label{fig:gnn_architecture}
\end{figure*}

\section{Introduction}
Simulations play a critical role in scientific discovery but are often computationally expensive, especially when studying complex phenomena like hydrodynamic instabilities. 
These investigations require exploring high-dimensional parameter spaces with large ensembles of simulations that generate terabytes of data. 
Generative surrogate models offer a potential solution to alleviate these computational demands: Once trained, they provide rapid predictions for new input parameters, enabling efficient design optimization and interactive analysis.
However, training these surrogate models requires large volumes of high-fidelity simulation data, placing significant demand on storage, memory, and I/O resources.

Although lossless compressors (e.g., gzip) often provide limited data reduction, lossy compression techniques such as ZFP~\cite{lindstrom2014fixed} and SZ~\cite {di2016fast} can achieve significantly greater compression while introducing errors that are visually or scientifically negligible.
Furthermore, incorporating lossy compression into the neural network training pipeline can mitigate resource constraints, but several questions remain: How much compression of the training data can be applied without compromising the quality of the final surrogate model?
Do compression errors propagate through the training pipeline and degrade the final model performance? Furthermore, does the use of compression add computational overhead during model training?

Compression ratios are often chosen arbitrarily. Users may err on the side of caution by either avoiding compression altogether or selecting conservative settings that limit the potential gains, or using overly aggressive compression that can degrade model quality. This ad hoc approach can either miss out on valuable efficiency gains or, more critically, compromise the fidelity of the surrogate model. 
In addition, users have primarily relied on broad, single-number statistical metrics, such as prediction accuracy or average error, to quantify the impact of compression errors during model training. 
While these metrics are straightforward to compute, they fail to identify model performance variation across different simulation outputs. 
Furthermore, metrics like prediction accuracy provide a limited understanding of the underlying physical properties of the simulations, which are crucial in scientific domains~\cite{laney2013assessing}. 
As a result, they fail to capture how compression truly affects the fidelity and usability of surrogate models, highlighting the need for more nuanced evaluation methodology. 

In this work, we present a novel approach, grounded in universal approximation theory~\cite{hornik1989multilayer}, to identify the compression error tolerance that is within the range of model-induced errors. Additionally, we utilize the inherent variability in neural network training to assess the impact of compression error. Training variability arises from factors such as random weight initialization, batch shuffling, and stochastic gradient updates, which can 
lead to different model outcomes under identical settings~\cite{zhuang2022randomness} with the same training data. 
By comparing the effects of compression error to the natural variance introduced by training randomness, we gain insight into whether compression introduces significant degradation. We propose that if the compression-induced error is less than the variability due to training, then compression can be considered benign.
We evaluate our method using both quantitative metrics (e.g., errors, physical properties) and qualitative visual assessments. Experimental results show that maintaining compression errors below the training variability allows surrogate models to maintain high accuracy while achieving substantial savings in data storage.
Furthermore, we demonstrate that lossy compression not only reduces storage requirements but also improves training performance by reducing training time by up to 3$\times$. The main contributions of this work are the following:
\begin{itemize}    

    \item We demonstrate that our approach enables the use of ZFP compression to achieve exceptional reductions in training data of 23.7$\times$ and 39.0$\times$, while maintaining high-fidelity surrogate models and reducing training time by up to 3$\times$.
    \item We propose a novel approach to leverage the variability in neural network training for evaluating the impact of lossy compression on surrogate model quality.
    \item We introduce an efficient method based on the universal approximation theorem to identify a compression error tolerance that preserves model accuracy.
    \item We perform an evaluation study across two scientific simulation benchmarks using both quantitative and qualitative assessments to validate our approach.
\end{itemize}

\begin{figure*}[ht]
\centering       
\includegraphics[width=\linewidth]{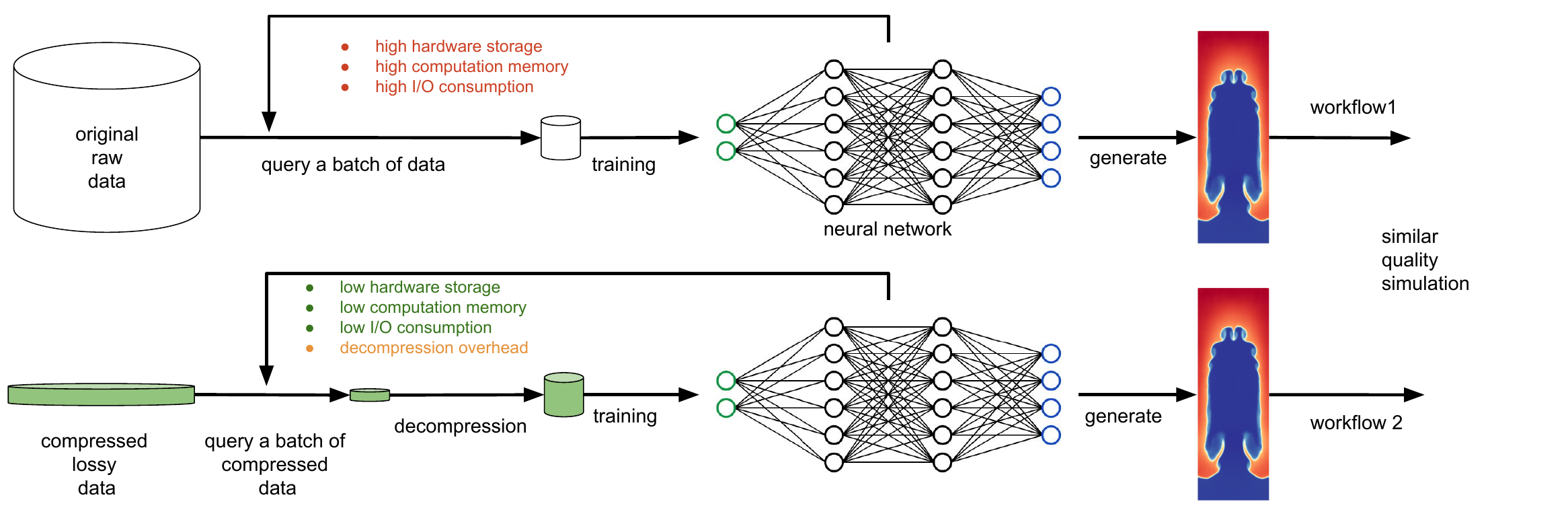}
    \caption{The workflow of training generative surrogate models with raw data (workflow 1) and lossy-compressed data (workflow 2).}
\label{fig:lossy_data_workflow}
\end{figure*}

\section{Background}
\label{sec:background}
To compare the effects of lossy compression and randomness in neural network training, it is crucial to create a fair and controlled environment. 
We conduct experiments on two representative simulation datasets---Rayleigh–Taylor and Richtmyer-Meshkov instabilities---to validate our observations and methodology, and to assess the quality of the final surrogate models using image quality and physics-based metrics.
The Rayleigh–Taylor (RT) instability occurs when fluids of different densities mix, as one fluid accelerates into the other. A common example is water suspended over oil within the Earth's gravity field. This process has broad scientific significance, contributing to the understanding of phenomena such as volcanic eruptions, supernovae, fusion energy, and more.
PCHIP~\cite{jekel2024machine} stands for piece-wise cubic Hermite interpolating polynomial, which were used as the initial geometry to seed a perturbation for a 
Richtmyer-Meshkov (RM) instability. RM and RT are closely related, and both instabilities occur in inertial confinement fusion (ICF) capsules \cite{zhou2025instabilities}. 

\textbf{Simulation datasets} are comprised of an ensemble of physics simulations. The ensemble is performed over multiple instances of parameter settings that govern, e.g., initial conditions and fluid properties. Each simulation result contains temporal solutions for multiple fields (e.g., density, energy), which are used to train the neural network model.
We assess the effect of lossy compression using two simulation benchmarks.
The corresponding data sizes and grid resolutions are summarized in Table~\ref{fig:simulation_data_table}.
The two benchmarks vary in their input parameters, but both output the same six fields: density, velocity ($x$ and $y$ components), pressure, energy, and material.
Training inputs are generated through uniform sampling across each parameter dimension. Each simulation result contains solutions at 51 time steps, where turbulent behavior increases with time. The training process considers each simulated time step as a separate sample. The RT simulation has 102,000 samples, and the PCHIP simulation has 152,184.

\textbf{Model architecture and training configuration} are critical to the performance of the surrogate model.
For model training, we apply convolution architectures, and the overall framework uses DCGAN~\cite{radford2015unsupervised} as the backbone for the generative surrogate model. 
However, our analysis is not constrained by this specific architecture and can be generalized to other widely used models, such as vision transformers~\cite{dosovitskiy2020image}.
Our architecture is shown in Fig.~\ref{fig:gnn_architecture}; it is a nine-layer convolutional neural network model.
In our study, we use the $L_1$ loss for model training, and the overall loss function can be calculated as follows:
\begin{equation}
    loss = \sum_{i=1}^{n}||\tilde{f}(x_i)- f(x_i)||_1
\end{equation}
The high-dimensional vector $x$ represents the input parameters of the network, and $n$ represents the total number of samples. $\tilde{f}(x)$ is the output of the surrogate model, which generally consists of several multi-dimensional, time-varying fields.
The model is trained with $f(x)$, which is the output of the simulation code, and the overall goal is to learn how to map the input parameters and the simulation outputs. 
$n$ is the total number of training ``samples,'' each of which corresponds to a vector of input parameters.
We use the Adam optimizer with a learning rate of 0.0001. 
\ZL{The dataset is split into $90\%$ simulations for training and $10\%$ for testing.}
The batch size for the Rayleigh-Taylor and PCHIP simulations is 64 and 16, respectively. \ZL{The selected number of epochs is 250 and 125, which ensures that models converge properly. The training setting and models used in this study have already been experimented with by researchers from mechanical engineering for their simulation study~\cite{jekel2024machine}.} The available GPU memory is sufficient to hold only a single batch during training. The training routine and model architecture are available at~\cite{professor}.

\begin{table}[t]
\centering
\tabcolsep 0.5em%
\caption{Data description of the Rayleigh–Taylor instability and PCHIP simulations.}
\begin{tabular}{cccc}
 \hline
Simulation  & Size  & Dimensions & \# of samples\\  \hline
  RT  &  450 GB &  $768 \times 256 \times 6$ & 102,000 \\  
 PCHIP    & 893 GB & $512 \times 512 \times 6$ & 152,184 \\\hline
\end{tabular}
\label{fig:simulation_data_table}
\vspace{-2mm}
\end{table}

\textbf{Quantitative metrics} measure the quality of the surrogate model output. 
This study includes \textbf{PSNR} (Peak Signal-to-Noise Ratio),  an objective image quality metric that measures the similarity between two simulations. 
It is often used to measure the quality of a distorted or compressed image. 
Complementing the image quality metric, physical metrics~\cite{laney2013assessing} provide an alternative perspective of simulation quality. \ZL{
In our study, we employ physics-based metrics, including conservation of \textbf{mass}, \textbf{momentum}, and \textbf{mixing layer thickness}, to characterize simulation properties.}
In the following equations, $\rho_i$ represents the (areal) density of the $i^\text{th}$ grid cell. $v_i^x$ and $v_i^y$ represent the velocity in the $x$ and $y$ direction.  $A$ is the area of a grid cell, which is constant over the domain. Note that the simulation code conserves both mass and momentum, but is subject to discretization error as the continuous domain is represented with several grid cells.

\begin{align}
m &= \sum_i A \rho_i & \text{Total mass} \\
p &= (p^x, p^y) = \sum_i A \rho_i v_i & \text{Total momentum}
\end{align}

\pl{%
We additionally evaluate model accuracy in terms of \emph{mixing layer thickness}, which is a common quantity of interest studied in the evolution of the Rayleigh-Taylor instability (see, e.g., \cite{laney2013assessing}).
We adopt the definition from~\cite{cook2004mixing}, which in our case reduces the time-varying mixing layer thickness to:
\begin{equation}
h(t) = H - \frac{2}{\rho_2-\rho_1} \int_{y} \left|\bar{\rho}(y, t) - \frac{\rho_1 + \rho_2}{2}\right| \mathit{dy},
\end{equation}
where $H$ is the height of the domain, $\rho_1 < \rho_2$ are the fluid densities, $\bar{\rho}(y, t)$ is the mean density at time $t$ over the horizontal slice at $y$, and the integration is approximated with summation over grid cells.  Note that $h(0) \approx 0$ and then generally increases over time as the two fluids start mixing.  To assess accuracy, we measure the correlation between time series $h(t)$ for surrogate models and the simulation ``ground truth.''
}

\textbf{Workflow} of the neural network training pipeline with raw and lossy-compressed data is illustrated in Fig.~\ref{fig:lossy_data_workflow}. The scale of modern simulation data is too large and full decompression may be infeasible; in such cases, training even requires online decompression to access the data efficiently, which places demands not only on I/O throughput but also on decompression performance.
During training, the model repeatedly retrieves batches of data for optimization.
Training and evaluation with raw data require substantial hardware storage and memory for the computation, and high I/O throughput.
In contrast, using compressed data for training reduces storage requirements, computation, and I/O load.
However, the process introduces additional decompression overhead during training. 
This study aims to understand the impact of introducing lossy compression on training efficiency and final model quality relative to the baseline that uses raw uncompressed data.

\section{Balancing the Model Variability and Lossy Compression Error}\label{sec:methods}

Prior work~\cite{poyser2021impact} has used lossy compression for classification tasks.
The performance of neural network models in classification is typically assessed by measuring prediction accuracy on a test dataset~\cite{poyser2021impact}.
Models exhibiting comparable accuracy on the test dataset are considered to be similar.
However, the evaluation methodology is particularly challenging to extend to the generative model, where assessing quality is inherently complex.
Researchers exploring generative surrogate models rely on qualitative evaluations, manually inspecting numerous output simulations to measure performance. 
The process, while insightful, is labor-intensive and time-consuming.
The classical approach, such as Fr\'{e}chet Inception Distance (FID)~\cite{heusel2017gans,sajjadi2018assessing}, requires a large number of samples to perform the comparison, but provides limited information about the underlying physical properties and visual quality of the generated output.

In this study, we perform quantitative measurements on physically relevant metrics (mass, momentum, and mixing layer thickness) and visual quality metrics (PSNR) to compare ground truth simulation outputs with model outputs.
For example, we use PSNR to measure the similarity between two simulations.
We record the PSNR distribution of a model output over the testing dataset to compare model quality.

\subsection{Comparing the Impact of Lossy Compression with the Variability in Neural Network Training}

The neural network training process involves inherent randomness that influences the final model's quality.
For example, model weights are typically initialized randomly~\cite{he2015delving}, and training data is shuffled in each epoch, leading to varied weight updates during optimization.
Additionally, some architectures incorporate random elements, such as dropout layers, which deactivate certain neurons to enhance generalization.
All the above factors introduce randomness into the model training process.
Therefore, models trained on the same dataset and hyperparameters can behave differently.

Determining whether differences between models stem from the randomness of the training process or lossy compression is critical to understanding the impact of compression errors.
If the training randomness causes greater variation between models than lossy compression does, compression-induced errors can be considered negligible.
However, if lossy compression introduces more error and its impact is greater than training randomness, then less error should be introduced.

\begin{figure}[t]
\centering   
 \includegraphics[width=0.9\linewidth]{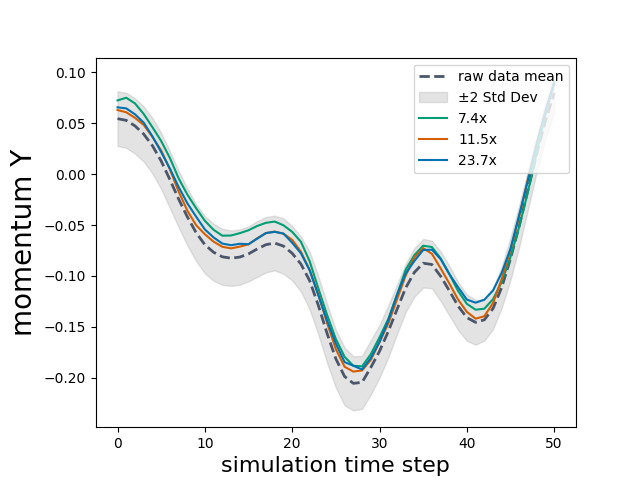}
\caption{Comparing total $y$-momentum vs.\ time between 30~models (gray band) trained on the same raw data and three models (colored curves) trained on lossy-compressed data with different degrees of loss associated with varying compression ratios (numbers in the legend).
The difference in momentum between raw data models and lossy models is qualitatively indistinguishable, indicating that the impact of the compression error is benign.}
\label{fig:lossy_psnr_evaluation_simulation_rt_randomness}
\end{figure}

To understand the randomness of trained models, we select a set of models trained with the same data and hyperparameters and compare their outputs with the ground truth simulation data.
Using the Rayleigh-Taylor instability simulation as an example, we compute $y$ momentum of each simulation's time step and plot the resulting curves in Fig.~\ref{fig:lossy_psnr_evaluation_simulation_rt_randomness}.
For the same simulation input, the dashed black curve represents the mean total $y$-momentum of each time step over 30 model outputs. The gray band represents $\pm 2\sigma$, which is the two standard deviation value range that covers $95\%$ confidence.
The gray band---30 models trained on uncompressed data, revealing that, even under identical settings, the output of neural-network surrogates can be different. 
The three colored curves (green, red, blue) represent three models trained on compressed data with varying degrees of loss.  As is evident, the gray band contains these curves. 
The model outputs generated from lossy and raw training data preserve physical properties such as momentum conservation relatively well, but with slight oscillation.
Meanwhile, from Fig.~\ref{fig:lossy_psnr_evaluation_simulation_rt_randomness}, the gray band and the colored curves are for all practical intents indistinguishable.

\section{A Model-Centric Approach to Determining the Compression Error Tolerance}
Determining the error tolerance for lossy compression is a persistent challenge in many fields. In the absence of a systematic framework, choosing an appropriate tolerance becomes a tricky balancing act. Practitioners are often forced to choose this arbitrarily. On one hand, a conservative choice preserves model quality, forgoing significant storage and compute gains. On the other hand, an aggressive choice may degrade the surrogate model's fidelity. To avoid this guesswork, one could use a brute-force, iterative process, where users repeatedly compress the data at different levels, train a model to convergence, and evaluate its final quality to see if the compression error degraded the model~\cite{underwood2024understanding}. 
This iterative cycle is computationally prohibitive, making it impractical for many applications and raising an important question: Can we establish an upper bound for the allowable compression error without resorting to expensive, repeated model training? 
\pagebreak

Our work is motivated by a key insight we developed in the context of generative surrogate models, which we describe here.
While some studies~\cite{sajjadi2018assessing, baker2016evaluating} have focused on finding an ideal tolerance that preserves the integrity of the data itself (Threshold~1 in Fig.~\ref{fig:compression_threshold}), our work takes a different approach tailored to generative surrogate models.
It is well established that neural networks exhibit high tolerance to errors, as demonstrated by their effective use of reduced precision and quantization techniques in training~\cite{nagel2021white}. 
More importantly, because of their finite capacity, models cannot capture every minute detail in the data. 
This limitation defines a second, more practical threshold: the precision level that retains all the features the model is capable of learning (Threshold~2 in Fig.~\ref{fig:compression_threshold}). Compressing data beyond this second threshold is detrimental, as it may cause the surrogate model to learn artifacts introduced by lossy compression. Our approach focuses on understanding and respecting this second threshold during compression to ensure a high-fidelity surrogate model.

\begin{figure}[t]
\centering       
\includegraphics[width=\linewidth]{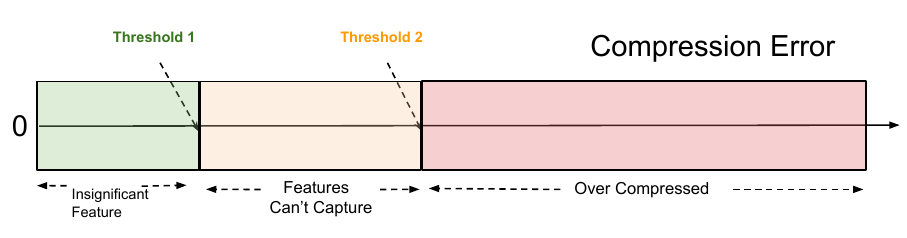}
    \caption{Two lossy-compression error thresholds for surrogate model training data. The first threshold represents the precision required to preserve the original data's integrity (e.g., the last few floating-point bits), while the second threshold corresponds to the level of detail that the surrogate model is able to learn. Compressing beyond the second threshold risks introducing artifacts that degrade model fidelity.}
\label{fig:compression_threshold}
\end{figure}



\subsection{A Neural Network is a Universal Approximator}
We propose a novel and efficient solution for setting the upper bound on compression error based on the Universal Approximator theorem~\cite{hornik1989multilayer}. 
The neural network is considered a universal approximator~\cite{hornik1989multilayer,zhou2020universality}, which means that with sufficient capacity (sufficient hidden units and layers), the network can approximate any complex function to a desired accuracy.

Let $y^{true}_i$ denote the high-fidelity simulation, $y^{lossy}_i$ the compressed training data, and $\hat{y}_i = f_\theta(x_i)$ the model output. 
According to the theorem, the model error $\epsilon^{model}_i = ||\hat{y}_i - y^{true}_i||$ should be near zero when trained on the ground-truth data $y^{true}_i$. 
When training on compressed data, the model error is expected to approximate the compression error $\epsilon^{loss}_i = ||y^{lossy}_i - y^{true}_{i}||$, since the model can only capture features present in the training data. In other words, training on lossily compressed data will result in $\epsilon^{model}_i \approx \epsilon^{loss}_i$. However, due to the model's limited capacity and the complexity of the training data, neural networks rarely attain this level of accuracy.

Our main hypothesis is that if the compression-induced error, $\epsilon^{loss}_i$, is lower than the inherent error of a model trained on lossless data, it will not degrade the model's performance. In other words, we want $\epsilon^{loss}_i \le \epsilon^{model}_i$. The rationale is that any information that the model is incapable of capturing can be removed from the training data via compression without degrading performance. In essence, the model's own prediction error on lossless data (Threshold~2 in Fig.~\ref{fig:compression_threshold}) provides a practical upper bound on allowable compression error.




To validate this hypothesis, we retrained a new surrogate model on the outputs from the original surrogate model trained on lossless simulation data and compared the results with those produced by the original model.
Fig.~\ref{fig:model_output_training} shows the distribution of the $L_1$ error between the model output and the simulation output.   
Note that we use the entire error distribution for this comparison, rather than a single aggregate metric, to capture sample-wise variations in error. The gray curve represents the error distribution of the model trained on the raw simulation output, while the blue curve shows errors for the (secondary) model trained on the output of the (primary) model. Their final $L_1$ error has a near-identical distribution.
In other words, this process of training on already-approximate data exhibits no significant ``generation loss.''
This result demonstrates that model output error can guide compression decisions. In our work, we use the model's $L_1$ error to define safe compression bounds (Threshold 2 in Fig.~\ref{fig:compression_threshold}), ensuring that compression errors remain within the model's error.

Algorithm~\ref{alg:find_desire_compression_ratio} outlines our method for finding the error tolerance for each sample.
\ZL{The main process is to iteratively double the tolerance on $L_\infty$ error as long as the observed $L_1$ compression error does not exceed the $L_1$ model output error $e$. We start our initial compression error tolerance guess with
$t = \frac{4^d e}{c(d)}$, where in our case $d = 2$ dimensions.
Based on the latest ZFP error distribution analysis~\cite[Appendix~A]{fox2026enhancing}, $c(2) \approx 1.089$ gives an expected $L_1$ error of $e$.
We then successively double $t$ as long as the observed $L_1$ compression error does not exceed $e$.
For the given initialization, this process empirically completes in 1 to 2 iterations.
The algorithm is performed under a per-sample setting and adaptively decides the desired compression ratio individually for each sample in the dataset.
During this iterative process, the compressed data only needs to refer to the error of the lossless model without training any additional models. }


\begin{algorithm}[t]
\SetAlgoLined
\KwResult{Compression error tolerance $t_i$ for each sample}
$S$ is the training sample space\;
$F$ is the surrogate model output\;
$Z$ is the output of compression/decompression\;
$i$ is the sample index\;

\ForEach{$s_i \in S$}{
    Initialize the error tolerance $t_i = \frac{4^d}{c(d)} ||f_i-s_i||$\\
    Compress using $t_i$ to produce $z_i$\\
    \While{$||z_i-s_i|| < ||f_i-s_i||$}{
       Double $t_i$ and update $z_i$ by recompressing
    }
}
\caption{Determining compression error tolerance}
\label{alg:find_desire_compression_ratio}
\end{algorithm}


\begin{figure}[!t]
\centering  
\includegraphics[width=\linewidth]{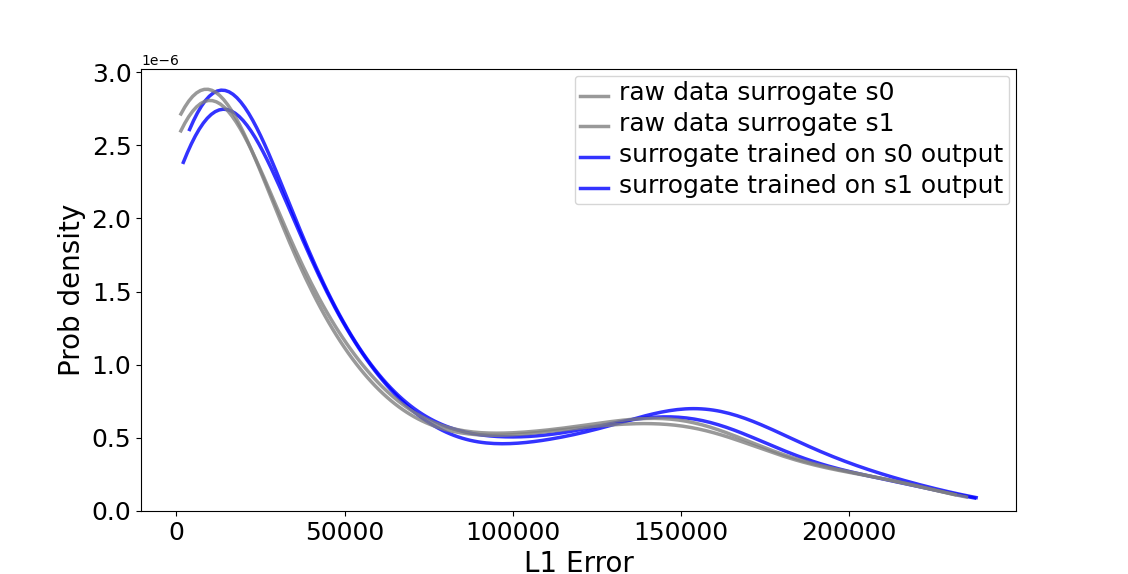}
\caption{Using model output to train a new model has a close $L_1$ error as the lossless model. This implies that the model output error captures the model capacity and can be used to guide the training data compression.}
\label{fig:model_output_training}
\end{figure}



\section{Quantifying the Impact of Data Reduction}
\begin{figure*}[hbt!]
\centering   
    \includegraphics[width=0.32\linewidth]{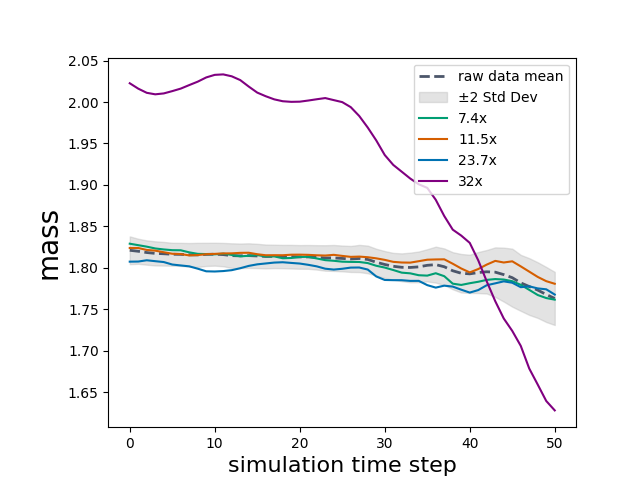}
    \includegraphics[width=0.32\linewidth]{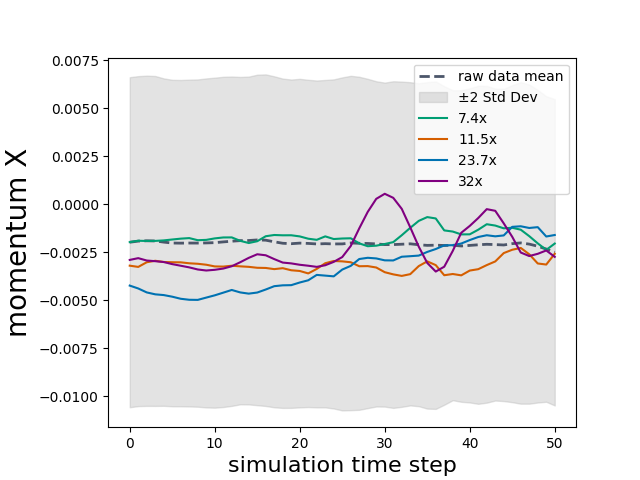}
    \includegraphics[width=0.32\linewidth]{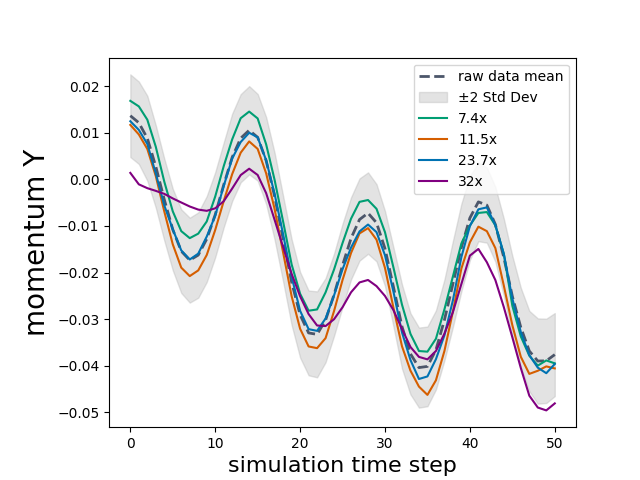}
    \caption{We calculate physical quantities to compare the output similarity of raw data models and lossy models in Rayleigh-Taylor (RT) simulations. The evaluation shows that the time-varying physical quantities associated with the surrogate model trained on lossy data up to 23.7x compression align closely with those of the raw data models.}
\label{fig:lossy_physical_evaluation_simulation_rt}
\end{figure*}

To validate the usability of our method, we conducted quantitative and qualitative evaluations to assess the impact of lossy compression during model training. We began by applying physical metrics and visual quality measurements to compare the lossy and lossless surrogate models. Next, we performed a qualitative analysis to demonstrate the effects of compression during model training.

\subsection{Quantitative Evaluation of Data Precision Reduction}
\ZL{We uniformly selected 200 simulations from each dataset as the test dataset, with the remaining simulations used for training.}
We trained five \textbf{raw data models} without data compression and compared them with \textbf{lossy models} trained on data that underwent lossy compression.
\ZL{For the Rayleigh-Taylor instability simulation, each model was trained for 250~epochs with a batch size of 64 and a learning rate of $1 \times 10^{-4}$. 
In the PCHIP experiments, models were trained for 125~epochs with a batch size of 16 and a learning rate of $5 \times 10^{-4}$.
The number of training epochs is chosen such that the loss function converges and shows minimal change after the designated epoch.}
The experiments were conducted on LLNL's Lassen supercomputer, whose nodes are equipped with 256~GB of main memory and four NVIDIA~V100 GPUs, each with 16~GB of graphics RAM.
Each model was trained using six nodes and 24~GPUs.
For ZFP data compression, we utilized the  \href{https://hdf5plugin.readthedocs.io/en/stable/}{hdf5plugin} Python module, which implements multiple lossy compression algorithms and integrates seamlessly with the PyTorch training environment.

\begin{figure*}[hbt!]
\centering   
    \includegraphics[width=0.32\linewidth]{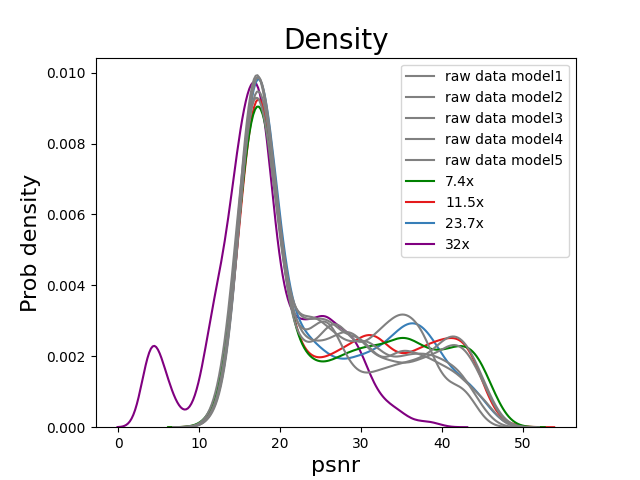}
    \includegraphics[width=0.32\linewidth]{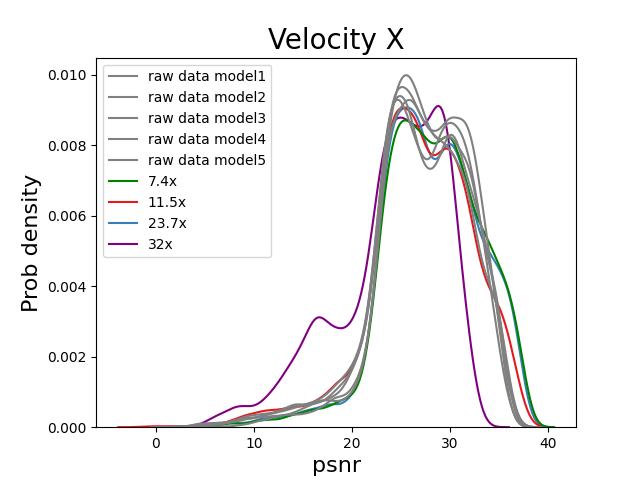}
    \includegraphics[width=0.32\linewidth]{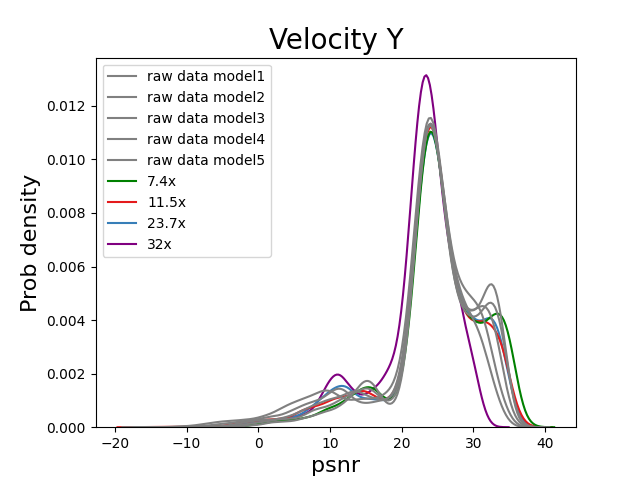}
    \includegraphics[width=0.32\linewidth]{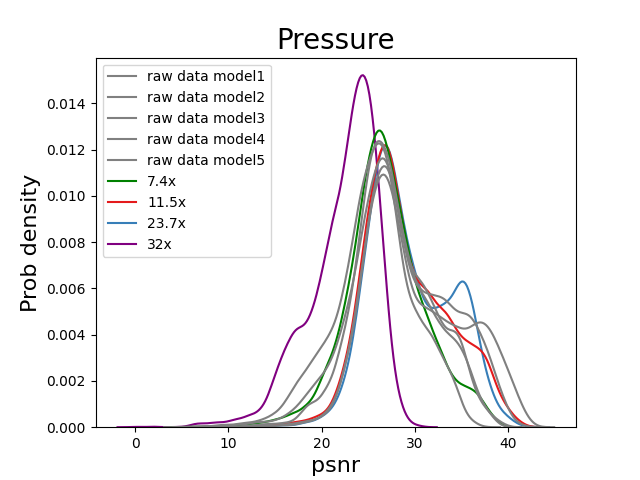}
    \includegraphics[width=0.32\linewidth]{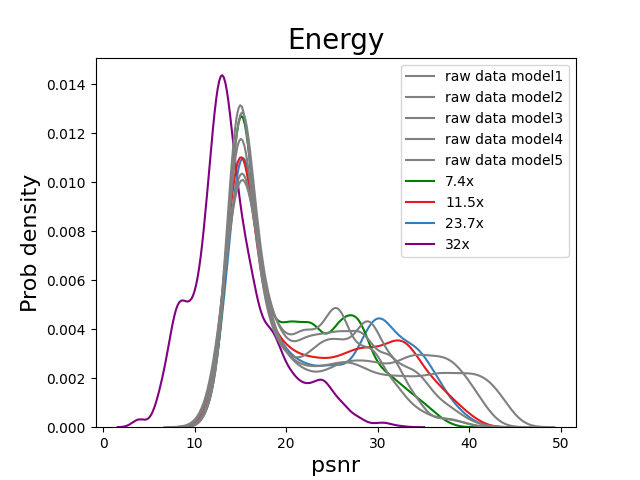}
    \includegraphics[width=0.32\linewidth]{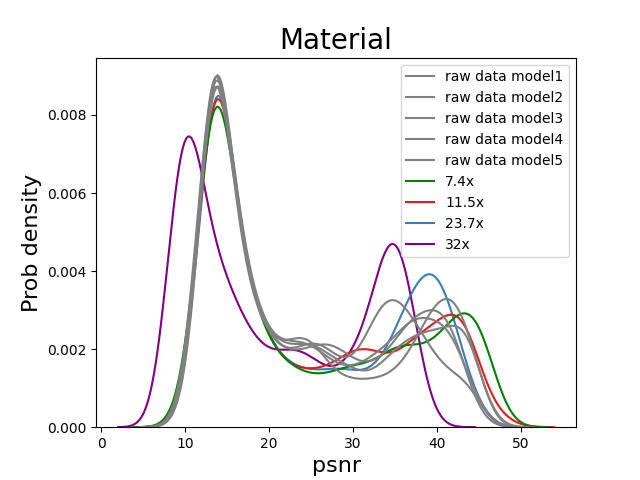}

    \caption{Using PSNR to compare lossless and lossy model output quality in the RT simulation. The evaluation indicates that the simulation can be compressed up to 23.7x without shifting the output PSNR distribution of the model.}
\label{fig:lossy_psnr_evaluation_simulation_rt}
\end{figure*}

\begin{figure}[t]
\centering   
 \includegraphics[width=0.9\linewidth]{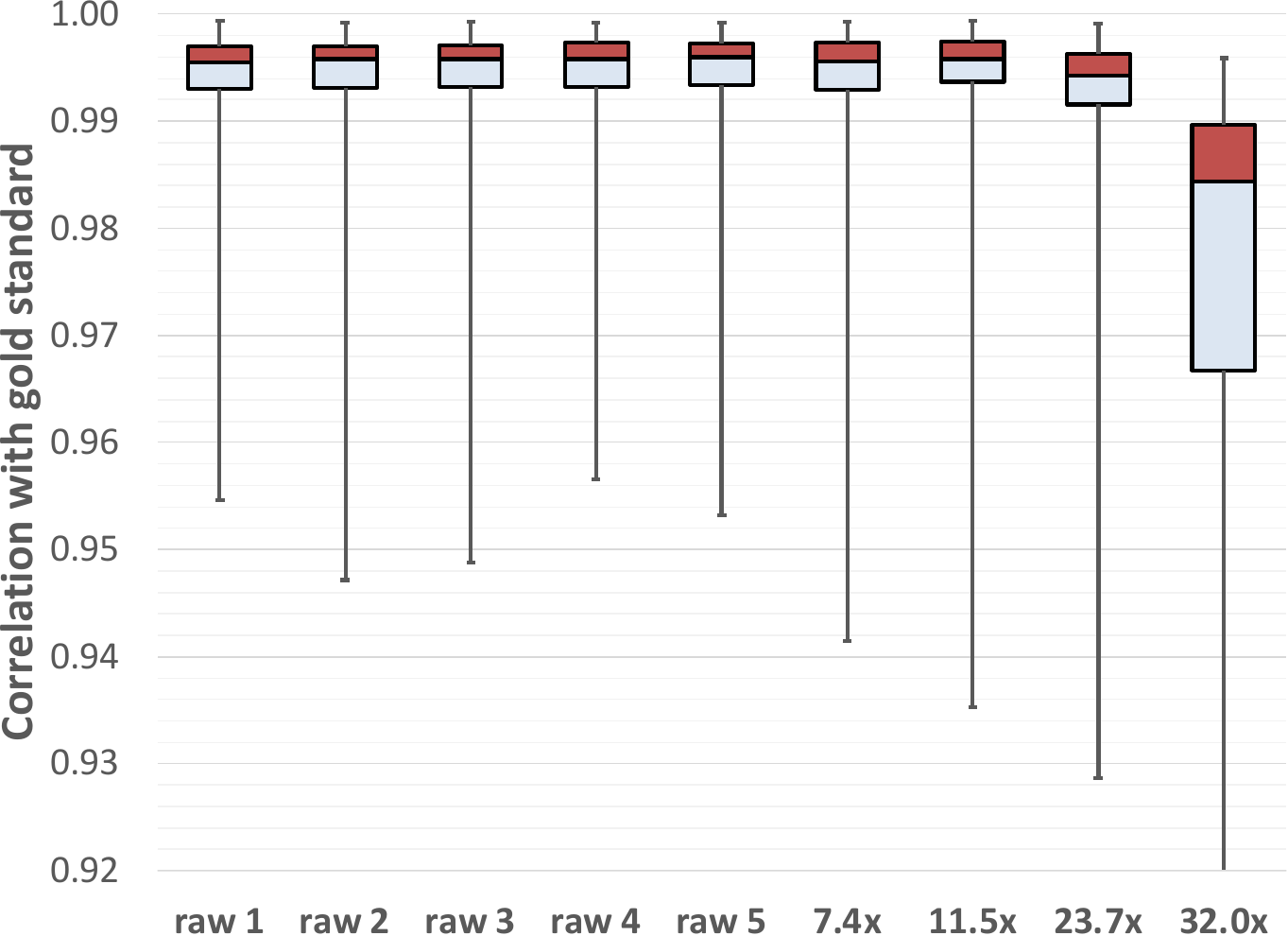}
\caption{\pl{Box plot of mixing layer thickness time series correlation between surrogate models and simulation data for 200 ensemble members.}}
\label{fig:mixing_layer_thickness}
\end{figure}

In Section~\ref{sec:methods}, we assessed the randomness of model training by comparing five raw data models with one another.
Our analysis utilized physical metrics such as total mass and momentum.
Fig.~\ref{fig:lossy_physical_evaluation_simulation_rt} illustrates the physical property values for a Rayleigh-Taylor (RT) simulation. 
It shows five gray curves representing the raw data models and three colored curves representing the lossy models.

These colored curves 7.4x green, 11.5x red, and 23.7x blue (the compression ratio is determined by the selected error tolerance which increasing by 10 starting with 0.001.), closely overlap the gray curves, indicating that no discernible differences were observed in the three physical quantities between the surrogate model trained on raw data (gray curve) and the surrogate model trained on lossy-compressed data reduced up to 23.7x. 
Here, the 23.7x compression is calculated through Algorithm 1.
As shown in the figure, the model trained on 32.0x lossy-compressed data exhibits differences in mass compared with the raw data models and the rest of the lossy models. However, the evaluation result shows that 32.0x has great deviations in the $x$-momentum and mass compared to the raw data models. Similar behavior can be observed from the mixing layer thickness metric through the box-plot in Fig.~\ref{fig:mixing_layer_thickness}, which displays the correlation distribution of multi-models' output with respect to ground truth.
Up to 23.7x compression, the lossy model maintains a relatively similar correlation distribution. With 32.0x compression, the quality of the lossy model has a significant distribution shift compared with these raw data models.

Beyond quantitative physical conservation properties such as mass and momentum, it is also essential to evaluate the visual quality of the resulting simulations. To this end, we computed the Peak Signal-to-Noise Ratio (PSNR) between the ground-truth simulation data and model-generated output to quantify reconstruction fidelity. The PSNR values were computed for all evaluation samples, and their distribution (i.e., a density plot of PSNR values) reflects how consistently the model preserves visual quality across the dataset.
The experimental results based on PSNR analysis (higher PSNR is better) are presented in Fig.~\ref{fig:lossy_psnr_evaluation_simulation_rt}.
The gray curves represent the PSNR distribution of the raw data models, while the colored curves display the PSNR distribution of lossy models.
Four lossy models are compared, with compression ratios 7.4x, 11.5x, 23.7x, and 32.0x.
In each figure, we compare the PSNR distribution of the lossy models with the raw data models to evaluate the impact of lossy compression.
The final result indicates that the PSNR distributions of raw data and lossy models (7.4x, 11.5x, and 23.7x) are generally indistinguishable.
32.0x compression is highly different from the raw data models, especially in the \textbf{pressure} field.
The density and material fields associated with compression ratios of 7.4x, 11.5x, and 23.7x are still sufficiently similar to ground truth.
We can tell that the training dataset can be compressed by up to 23.7x from the PSNR distribution comparison.
With a compression ratio of 32.0x, the overall PSNR distribution shifts and can be easily distinguished from the other models.
This observation aligns with Fig.~\ref{fig:lossy_physical_evaluation_simulation_rt}.

Similarly, we experimented with the PCHIP simulation, with comparisons presented in Fig.~\ref{fig:lossy_evaluation_prsn_simulation_pchip}.
In this experiment, we compressed the training data with compression ratios ranging from 8x to 39x and 
evaluated the performance of the resulting lossy models with the lossless model.
We also perform a similar comparison with the PSNR metric value distribution.
In all fields, we are not able to distinguish the difference in the PSNR distribution between the lossless and lossy models. 
The final result indicates that under the PSNR measurement, we can compress the training dataset by up to 39x without significantly impacting the quality of the model output.

Some fields from the generative model are of low quality when compared with the ground truth simulation.
However, in discussions with domain experts, we learned that the model does not have to be perfectly accurate to be useful. Domain experts are willing to accept errors in the surrogate model as long as the model captures general physical behaviors and supports rapid exploration of input parameters.
This work mainly focuses on evaluating the impact of lossy compression on the generative surrogate model training and the output model quality.
Improving the quality of the surrogate model to generate more accurate simulation results is important, but out of the scope of this study.

\begin{figure*}[t]
\centering   
    \includegraphics[width=0.32\linewidth]{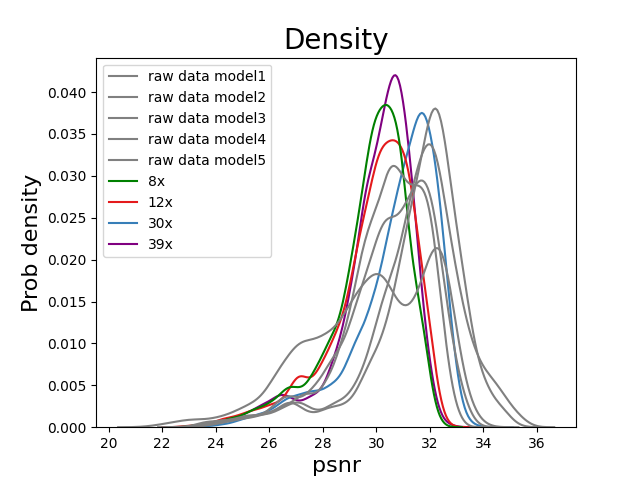}
    \includegraphics[width=0.32\linewidth]{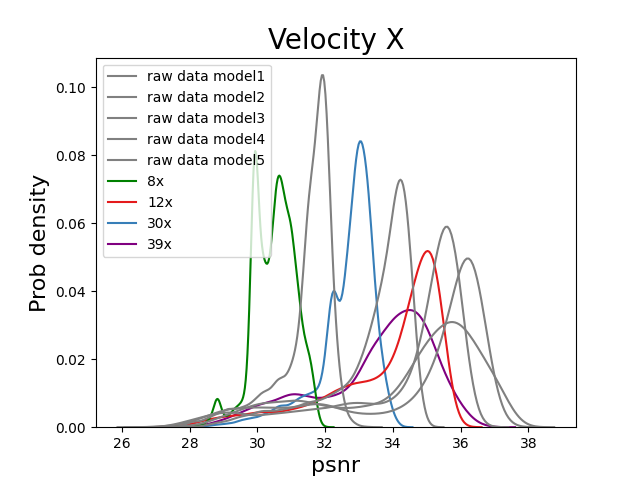}
    \includegraphics[width=0.32\linewidth]{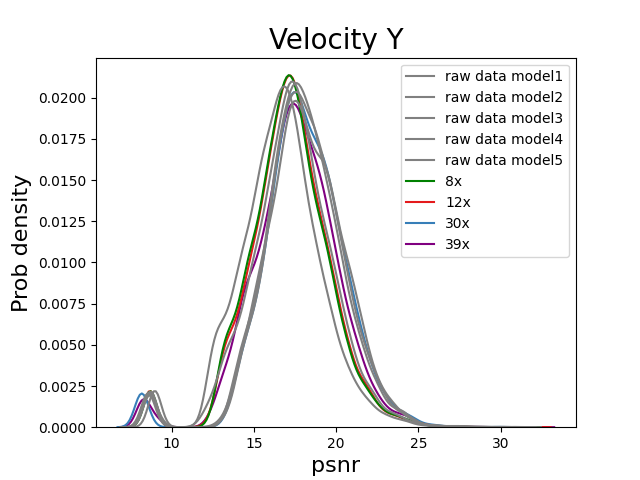}
    \caption{Using PSNR to compare lossy and raw data model output quality with the PCHIP simulation. The final evaluation demonstrates that we can compress the training data up to 39x without noticeable differences from the raw data models.}
\label{fig:lossy_evaluation_prsn_simulation_pchip}
\end{figure*}

\subsection{Qualitative Evaluation}
Our quantitative study evaluated the impact of lossy compression on model accuracy.
For the Rayleigh-Taylor simulation, the training data can be compressed up to 23.7x without decreasing model accuracy with respect to PSNR while also preserving physical conservation properties such as mass and momentum.
For the PCHIP simulation, the training data can be compressed up to 39x without significantly impacting PSNR distribution.
In addition to the quantitative evaluation, we also assess the models by qualitatively comparing the impact of lossy compression on the models trained with 23.7x (RT) and 39x (PCHIP) lossy-compressed data.
Instead of using specific metrics, we directly visualize these results to compare the differences between the lossy and lossless models.

Fig.~\ref{fig:qualitative_rt_comparison} presents a visualization of three fields for the RT simulation (\ZL{For display space considerations, the visualizations are rotated by 90 degrees}.)
In this comparison, we annotate the ground-truth simulation, the simulation generated by the lossless surrogate model, the 23.7x lossy model, and a model trained on 32.0x-compressed data.
The ground truth simulation shows more detailed features (e.g., edges, texture) than the simulations generated by the surrogate models.
Nevertheless, the neural network models successfully capture the primary features of the simulation.
When comparing the outputs generated by the lossless model and the 23.7x lossy model, it is still difficult to visually distinguish the differences between them.
Once the data is over-compressed to 32.0x, the quality of the surrogate model is also impacted. The model starts to produce outputs with missing features or blurs out the entire field. 


\begin{figure*}[t]
\centering   
    \includegraphics[width=\linewidth]{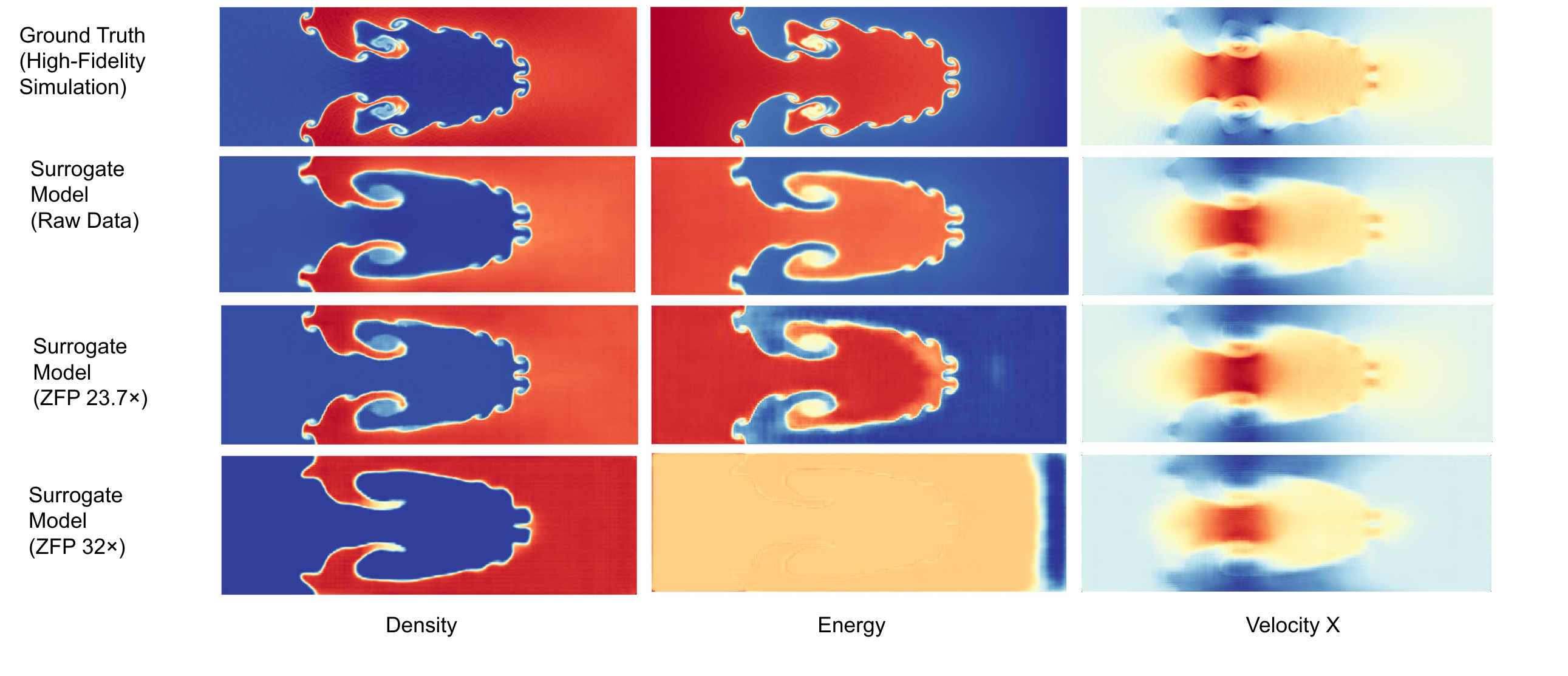}
    \vspace{-8mm}
    \caption{A visual quality comparison between a model trained on raw data and two models trained on lossy-compressed data.
    Compared with the ground truth simulation, the top two surrogate models capture the major geometric features of the simulation.
    However, at 32.0x compression (bottom), the resulting model loses significant features, especially in the energy field.}

\label{fig:qualitative_rt_comparison}
\end{figure*}

\section{Performance Comparison and Evaluation}
Beyond the compression ratio and the impact of lossy compression on model quality, another important consideration is the decompression overhead introduced during neural network model training when using compression.
In our study, the training pipeline performs online decompression during the neural network training process, as illustrated in Fig.~\ref{fig:lossy_data_workflow}.
While training with compressed data online can improve data loading performance and enable training on large datasets on storage- and I/O-limited devices, it also introduces decompression overhead that may reduce the overall I/O throughput.
During training, the neural network processes all samples once per epoch. 
In a distributed computation framework, after each epoch, the data gets randomly shuffled and redistributed across the compute nodes. 
Random data shuffling~\cite{meng2019convergence} is a standard practice in parallel neural network training, which is crucial for ensuring model convergence and preventing overfitting.
With random data shuffling, the decompression occurs each time a sample is accessed during training.
Evaluating the impact of this approach is essential for validating the practical utility of lossy compression.
To perform the training performance evaluation, we use the Rayleigh–Taylor (RT) instability dataset to measure the per-batch data loading throughput and per-epoch training time. The per-batch data loading time is the time it takes to load a batch of data for training, i.e., without model optimization steps.  The per-epoch training time is the time that models go through the whole dataset with model optimization operations.

\subsection{Data Loading Throughput during Training}
We conducted experiments on the model training process using various compression ratios, different numbers of GPUs, and three distinct file systems that are used by domain scientists to store their data. This study evaluated training performance under ZFP compression across three data storage file systems in a supercomputer environment.
Our experiment design measures data loading throughput within the PyTorch environment with the built-in data loader. 
For each experiment, we run the model for 30 epochs to measure the related data loading and training performance. 
Our experiment involved six nodes, each with four GPUs, of the Lassen supercomputer. The data loading performance measurements are performed over 24 GPUs total, and the final per-batch data loading throughput result is reported as an average value over the 24 GPUs.
This setting is the standard setup that domain experts use to train their models. To establish a baseline, we measure the per-batch data throughput and training performance for raw data training across three file systems, and the result is reported in Fig.~\ref{fig:rt_io_performance}. The per-batch data throughput measures how fast a batch of data is ready for the neural network, which excludes the weight update and back propagation steps during training.

\ZL{File System 1-\textit{workspace} is a network file system that is not specifically designed for parallel applications. File System 2-\textit{VAST}, and File System 3-\textit{GPFS} are high-performance parallel file systems; the study by Kogiou et al.~\cite{kogiou2023characterization} compares their I/O performance.
VAST has 4.83~GB/s write and 6.88~GB/s read bandwidth, and GPFS has 5.69~GB/s write and 10.24~GB/s read bandwidth. These numbers are reported as a mean value over multiple runs in their experiments.
A detailed I/O bandwidth of read/write comparison between VAST and GPFS with various numbers of nodes can be found in Kogiou's study~\cite{kogiou2023characterization}.}

Using one HDF5 file per sample without compression, File System 1 records the lowest per-batch data throughput (145.65~MB/s), followed by File System 2 (227.31~MB/s), while File System~3 reaches the highest at 746.7~MB/s. The per-batch data throughput is much slower than pure I/O, because additional operations, such as collation of data into a batch, memory allocation, and more, are required in the PyTorch training environment.
Follow-up experiments assessed the throughput across different compression ratios on these three file systems.
The throughput is then compared with the raw data performance. 
Under ZFP lossy compression, the per-batch data loading throughput across different file systems remained relatively consistent with minor variance.
The maximum throughput difference between raw data and lossy data on File System~1 is improved up to 2.96x, and on File System~2, the throughput is increased by 1.89x. In both systems, the data loading performance is improved using ZFP compression because less data is being transferred. 
However, on the much faster File System~3, the throughput of loading raw (uncompressed) data is higher.
Here, raw data throughput exceeds the serial decompression throughput supported by ZFP and HDF5's compression plugin interface.

\begin{figure}[t]
\centering   
    \includegraphics[width=\linewidth]{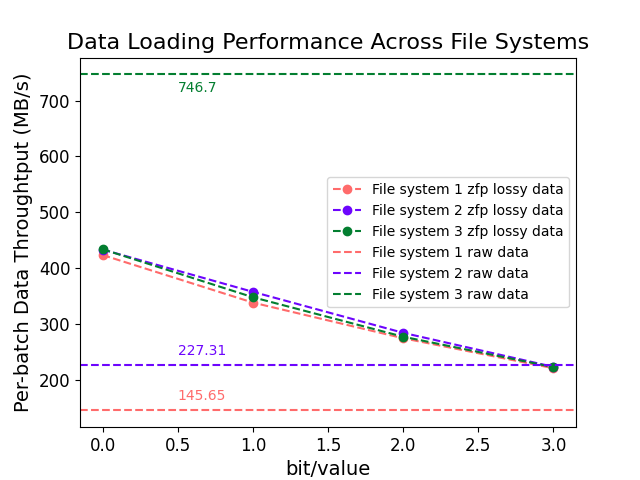}
    \caption{\ZL{The three file systems demonstrate varying per-batch data loading throughput when training on raw data. Except for File System~3, model training with lossy compression yields better data-loading performance than raw data on File System 1 and competitive or better performance on File System 2. The throughput of reading ZFP compressed data under different compression ratios is relatively similar across the three file systems.}}
\label{fig:rt_io_performance}
\end{figure}

\subsection{Model training performance Comparison}
Even though the data loading performance is different across the three file systems, it is important to look at overall timings as data loading is just one piece of the training workflow. To examine this difference, we extended our analysis to measure the total time required for the neural network to complete one full pass through the dataset, referred to as an epoch.
This measurement includes all operations involved in training, including data loading, decompression, model updates, and others. 
Similar to the previous experiments, we report the average per-epoch time, representing the time for processing the entire dataset once, averaged over 30 training epochs.

\begin{figure*}[!t]
\centering   
    \includegraphics[width=0.32\linewidth]{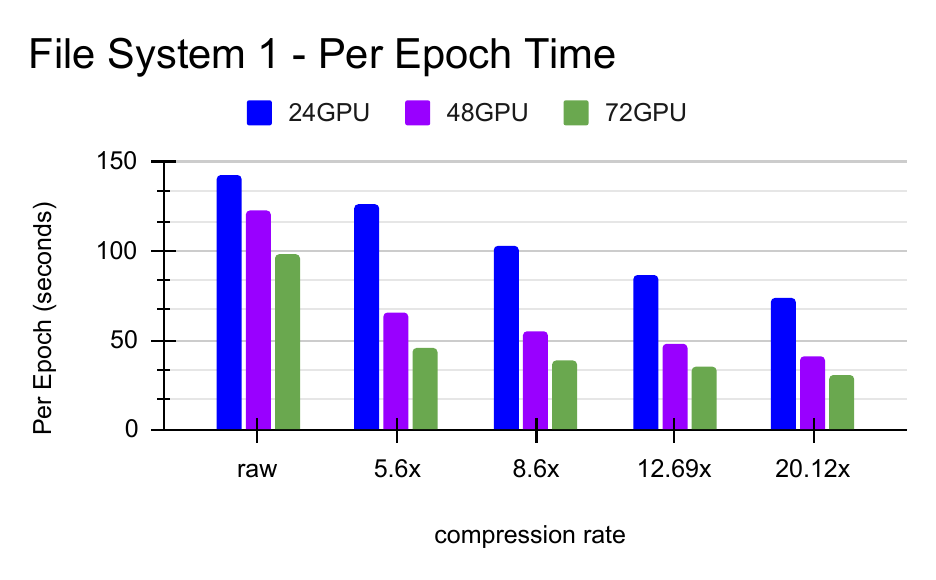}
    \includegraphics[width=0.32\linewidth]{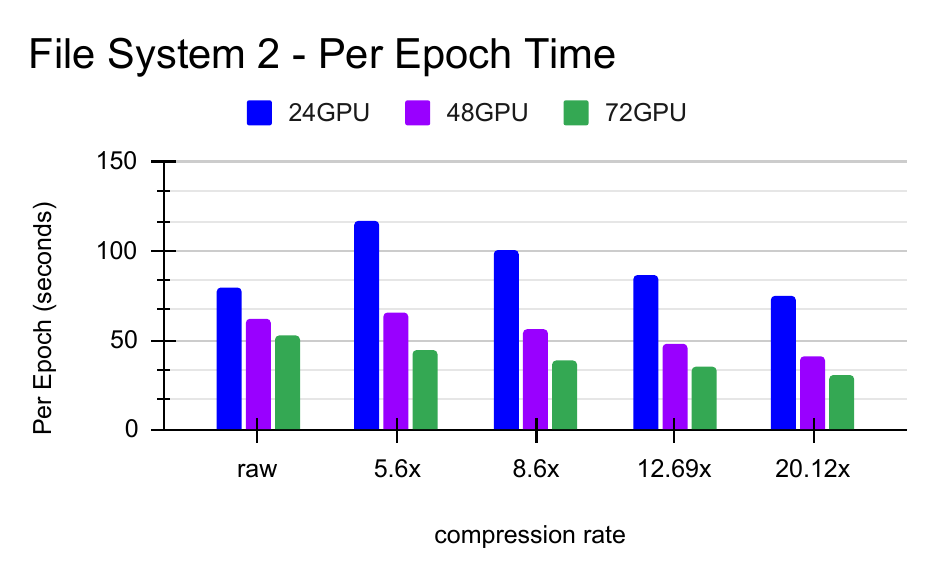}
    \includegraphics[width=0.32\linewidth]{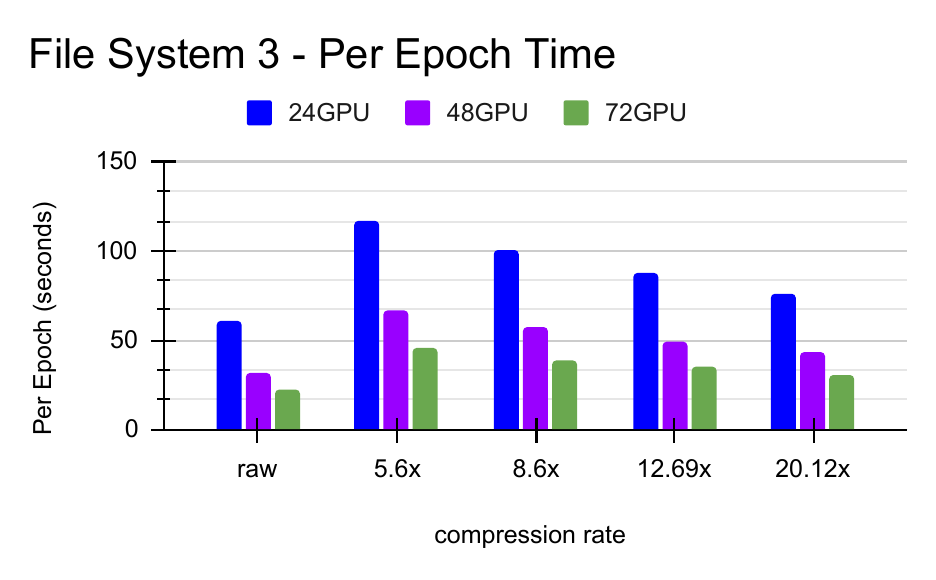}
    \caption{We measure the per-epoch time with 24, 48, and 72 GPUs. With the doubling of GPUs in training, we would expect to see the per-epoch time decrease close to half. The high I/O overhead of reading raw data precludes such linear speedup for File Systems~1 and~2.}
\label{fig:rt_training_performance}
\end{figure*}

The evaluation result is shown in Fig.~\ref{fig:rt_training_performance}.
It shows the amount of time that a model goes through all the training data and updates the model, and also provides information about how long it takes to train a model. 
In File System~1, taking the raw data as an example, with an increase in the number of GPUs, the time required to finish one epoch of training decreases with the increase in the number of GPUs. This trend can also be observed in the other two file systems.

On File System~1, the per-epoch time for compressed data is lower than for raw data, regardless of how many GPUs are used for training. With lower bit rates, the per-epoch time decreases and throughput increases. 
Lossy compression benefits the model training time on File System~1.
The per-epoch time for 20.12x lossy compressed data with 72 GPUs is three times less than for raw data. This observation is consistent with the per-batch data loading throughput trends shown in Fig.~\ref{fig:rt_io_performance}.
Compared with raw data and 5.6x lossy compressed data, the per-epoch time for 72 GPUs is reduced by half, but the per-epoch time for 24 GPUs drops only from 145 seconds to 125 seconds.
The same observation can also be made for File System~2. 
File system~2 shows that, when using 24 GPUs, training on raw rather than compressed data is faster. However, when using 48 and 72 GPUs, compression can benefit overall performance.
Comparing the highest compression ratio of 20.12x with raw data, the per-epoch data loading time is improved when training on compressed instead of raw data with 72 GPUs. This observation indicates that an increasing number of GPUs (thus reducing compute time) will increase the benefit of training with compressed data. 
On File System~3, a similar trend is demonstrated; increasing the compression ratio decreases per-epoch data loading time.  However, the best performance is achieved using raw data, as decompression time dominates the overall data loading time. 

Overall, this experiment demonstrates that ZFP lossy compression can not only reduce data storage but can also help reduce training time.
The decompression overhead during online training with compressed data can be balanced out by the reduced data loading time, which helps reduce the per-epoch training time. Using compression, the per-epoch training time is reduced over reading raw data on File Systems~1 and~2. The relative benefit of compression increases with the number of GPUs used to train the model.
\section{Related Work}

\subsection{Randomness of Neural Network Training}
Training neural networks in large-scale, distributed settings inevitably introduces randomness in the model output. 
Sources of nondeterminism include data partitioning~\cite{sogaard2020we}, stochastic optimization procedures, stochastic layers such as dropout, and hardware-level variability~\cite{zhuang2022randomness}.
As a result, models trained on the same dataset, hyperparameters, and optimizers can still produce markedly different predictions~\cite{summers2021nondeterminism}.
Prior work has analyzed the impact of randomness on replicating classification models~\cite{zhuang2022randomness}, while Wang and Jia \cite{wang2023data} showed that identical training configurations can yield divergent data-value metrics such as Shapley values \cite{ghorbani2019data} and leave-one-out errors \cite{bachmann2022generalization}. In this study, we extend these investigations to generative surrogate models and exploit the intrinsic randomness of model training to balance the compression error during training.


\subsection{Lossy Compression}
Lossless compression often provides limited reduction of scientific datasets. 
However, previous research has found that not every bit of data is required to answer scientific questions~\cite{hoang2020efficient,hoang2018study}.
Lossy compression~\cite{lindstrom2014fixed, di2016fast} is an approach that provides a much higher compression ratio than lossless compression without necessarily affecting the accuracy of subsequent data analyses. 
Poyser et al.~\cite{poyser2021impact} have applied lossy compression, such as JPEG image compression and video compression, to improve the efficiency of neural network training by reducing the transmission volume and data storage. 
This process introduces little performance loss in the final model prediction.
Jian et al.~\cite{jin2021comet} evaluate the impact of SZ compression on classification tasks and show that lossy compression helps improve memory management.

Previous studies have mainly focused on classification benchmarks such as ImageNet, where straightforward quantitative metrics (e.g., top-1 accuracy) make it easy to judge whether lossy compression affects model performance.
Generative surrogate models, in contrast, are harder to evaluate. 
The key question is not simply whether a prediction is correct, but whether the generated image or simulation outputs preserve the semantics of the target physical phenomenon. 
For fluid-dynamics data such as the Rayleigh-Taylor instability, the synthesized fields must remain physically plausible so that researchers can still interrogate the underlying flow behavior.
To this end, our work adopts established image-quality measures: peak signal-to-noise ratio (PSNR)~\cite{huynh2008scope}, and physics metrics (e.g., total mass, momentum) to compare surrogate model outputs with their simulation code-generated counterparts. These metrics enable a better assessment of how lossy compression and other data-reduction techniques affect the accuracy of generative surrogate models.


\subsection{Generative Surrogate Modeling for Scientific Simulation}
Scientific simulation plays a critical role in many scientific disciplines, but the computational cost associated with running these simulations remains a significant barrier.
The often high-dimensional space of simulation input parameters, which requires a large number of samples to thoroughly explore, further exacerbates this problem.
Generative surrogate models for the far more computationally expensive simulations are becoming a popular tool for researchers to mitigate computational cost. 
Instead of running expensive computations, researchers can train a generative surrogate model~\cite{hazarika2019nnva,he2019insitunet,shi2022gnn} to serve as a proxy for the simulation for fast analysis and parameter exploration.
Shi et al.~\cite{shi2022vdl} proposed a VDL-surrogate framework that trains an auto-encoder that encodes simulation fields as a latent representation. It then uses the latent representation to train a surrogate model for parameter exploration.
The work mitigates the training cost for the surrogate model but does not address the large data storage problem, and the same problem remains in training an autoencoder.

\section{Discussion and Future Work}
We have introduced the benefit of applying lossy compression to training data for generative surrogate model training and evaluated its impact. We designed an algorithm to efficiently search for the desired compression ratio without the need to repeatedly retrain. The proposed approach in this study can also be applied to compression algorithms other than ZFP. Here, we discuss the limitations of our approach. 

\textbf{Decompression Overhead:} Our experiments demonstrate that training on lossy-compressed data has the potential to benefit from reduced data loading overhead. When the file system already provides high throughput (File System~3), the computational overhead associated with decompression can potentially increase training time.
Compression plugins supported through HDF5-based data loading are currently limited to serial CPU execution and do not take advantage of the parallelism supported by ZFP and similar compressors. 
Specifically, our current implementation does not fully exploit available computational resources such as OpenMP or GPU acceleration. Therefore, the decompression overhead could be further mitigated through parallel decompression strategies leveraging multi-threading (OpenMP) or GPU-based implementations.

\textbf{Proxy Model for Universal Approximator:}
Determining the desired compression ratio for lossy compression through the universal approximator requires a trained model, which can be expensive. A potential direction to address this concern is to use a proxy model trained on a smaller subset of data, for example, using 10\% selected uniformly. Although such a model tends to overfit the subset, it can still provide useful estimates of acceptable compression levels. We could also estimate the compression ratio for the remaining data based on its similarity to the selected subset.

It is also important to note that the achievable compression ratio of the training data depends on the performance of the neural network model. 
High-performing models can achieve greater prediction accuracy but are more tolerant of errors in the training data, whereas lower-performing models can tolerate larger compression-induced errors.
\ZL{In the current research setting, our experiments focus on the DCGAN model~\cite{radford2015unsupervised}.
In the future, we will be interested in evaluating our approach on advanced models, such as the Vision Transformer, to see how our approach behaves in this kind of model setting.}

\section{Conclusion}
In this study, we investigate how lossy compression applied to training data impacts the training time and accuracy of a generative surrogate model.
Our findings reveal that neural networks trained on identical configurations can produce significantly different generative surrogate models due to random factors involved in the model training process, such as stochastic optimization, randomized data shuffling, and model weight initialization. Our study shows that, at an appropriate compression level, the error introduced by lossy compression has a smaller and often negligible effect compared to the inherent randomness in neural network training. 
We proposed a novel algorithm to calculate the desired compression error tolerance for training without affecting the quality of the final trained model.
Our experiments demonstrate that simulation data can be compressed up to 23.7$\times$ and 39$\times$ in two evaluation benchmarks without negatively impacting the quality of the generative surrogate model.
The use of lossy compression generally improves data loading throughput, resulting in as much as a 3$\times$ reduction in total training time.

\pl{
\section*{Acknowledgment}
This work was in part performed under the auspices of the U.S.\ Department of Energy by Lawrence Livermore National Laboratory under Contract DE-AC52-07NA27344. This research was also supported by the U.S. Department of Energy under grant DE-SC0023320. The manuscript is reviewed and released under LLNL-CONF-2007282.
}

\newpage
\IEEEtriggeratref{19}
\bibliographystyle{IEEEtran}
\bibliography{library.bib}
\newpage

\end{document}